\documentclass[aps,prb,twocolumn,showpacs,superscriptaddress]{revtex4-1}  % for review and submission
\usepackage{graphicx}  % needed for figures
\usepackage{dcolumn}   % needed for some tables
\usepackage{bm}        % for math
\usepackage{amssymb}   % for math
\usepackage{multirow}
\usepackage{amsmath}
\usepackage{natbib}
\usepackage{xcolor}
\begin{document}

\title{Intrinsic interface states in InAs-AlSb heterostructures}

\newcommand{\affA}{CNRS-Laboratoire de Photonique et de Nanostructures, route de Nozay, F-91460, Marcoussis, France}
\newcommand{\affB}{Laboratoire de Physico-chimie des Microstructures et Micro-syst\`emes, Institut Pr\'eparatoire aux Etudes Scientifiques et Techniques, BP51, 2070 La Marsa, Tunisia}
\newcommand{\affC}{FOTON, Universit\'e Europ\'eenne de Bretagne, INSA-Rennes and CNRS, Rennes, France}
\newcommand{\affD}{Tyndall National Institute, Lee Maltings, Dyke Parade, Cork, Ireland}
\newcommand{\affE}{Ioffe Physical-Technical Institute of the Russian Academy of Sciences, St. Petersburg 194021, Russia}
\author{F.~Raouafi}
\affiliation{\affB}
\author{R.~Benchamekh}\affiliation{\affA}\affiliation{\affD}
\author{M.O.~Nestoklon}
\affiliation{\affA}\affiliation{\affE}
\author{J-M.~Jancu}
\affiliation{\affC}
\author{P.~Voisin}
\affiliation{\affA}

       % D0 authors (remove the first 3 lines
                             % of this file prior to submission, they
                             % contain a time stamp for the authorlist)
                             % (includes institutions and visitors)
\date{\today}
\begin{abstract}
We examine the formation of intrinsic interface states bound to the plane of In-Sb chemical bonds at InAs/AlSb interfaces. Careful parameterization of the bulk materials in the frame of the extended basis $spds^*$ tight-binding model and recent progress in predictions of band offsets severely limit the span of tight-binding parameters describing this system. We find that a heavy-hole like interface state bound to the plane of In-Sb bonds exists for a large range of values of the InSb/InAs band offset.
\end{abstract}

\maketitle

\section{Introduction}

Ever since the seminal papers of I.Tamm\cite{Tamm}, the possible existence of intrinsic surface or interface states in semiconductors has been a hotly debated issue, but the emerging topic of topological insulators has recently renewed the interest in this field\cite{RPM_TI}, in connection with predicted ``quantum immunity'' of edge-state currents against scattering.  In the early 80's, a first type of interface state was predicted to occur in HgTe/CdTe heterostructures\cite{BastardHgTe,Chang,Schulman}, due to the boundary conditions between the inverted band structure of HgTe and the normal band structure of CdTe. 

In that case, existence of interface states is primarily due to anomalous properties of one of the bulk constituents. Much more recently, the attention was drawn to the original situation of interfaces between materials sharing no common atom (NCA), like InAs/GaSb, InAs/AlSb, BeTe/ZnSe or (InGa)As-InP. In these materials, interfaces involve chemical bonds that do not exist in the host materials, for instance In-Sb or Ga-As in the first example. Such interface bonds can in principle act as a local potential well that may capture the carriers. This second type of interface state, if it exists, relies mainly on local interface properties and proper modeling requires detailed atomic-scale information that is normally missed by the standard envelope-function approximation \cite{Bastard81,Bastard_book}.
The existence of interface states in the InAs/AlSb system was first suggested heuristically by Kroemer et al. in 1992 \cite{Brar} as a possible explanation for the heavy n-type doping observed in nominally undoped superlattices, but it was soon argued that possible values of band offsets would not allow for an interface state resonant with conduction band \cite{Dandrea}. Later, this question was revisited using ab initio methods by Shaw et al.\cite{Shawetal, Shaw}, who concluded to localization of the hole ground state near the plane of In-Sb bonds. However, in the bare DFT without spin-orbit interaction used in these calculations, InAs and InSb are metals rather than semiconductors \cite{Kim}, and this makes comparison to experimental results difficult. InAs/AlSb has a type II band line-up with ground electrons (holes) in the InAs (AlSb) layers, and reported values of valence band offset in the 100-200 meV range \cite{Ram-mohan, Kroemer_review}. In recent years, the InAs/AlSb system has proved its technological importance with the emergence of high performance optoelectronic devices based on intersubband transitions, such as quantum cascade lasers \cite{Teissier} in the mid-infrared. A need to better analyze and control interface composition was evidenced \cite{Ponchet}, as large strain can accumulate and lead to plastic relaxation when preferential formation of Al-As interface bonds prevails. Besides recent progress in electron microscopy that allow for chemical and strain analysis with sub-nanometer resolution \cite{Ponchet,Colliex}, it has become possible to observe directly the wavefunctions of quantum states using STM \cite{STM}. Thus, advances in fine material characterization offer a unique opportunity to get conclusive experimental proofs on the existence of intrinsic interface states. On the modeling side, progress in computing would now allow for ab-initio studies combining full atomic relaxation and realistic electronic structure of narrow gap semiconductors, but there is also room for computationally easier, yet predictive calculations based on empirical-parameter atomistic theories such as advanced empirical tight-binding (ETB) \cite{Jancu98} or atomistic empirical pseudo-potentials (AEPP)\cite{Zunger96,Bester09} schemes. A $sp^3$ ETB model with first and second neighbor interactions was first used by Theodorou et al. \cite{Theodorou}. Late, modeling of InAs/AlSb and ZnSe/BeTe within the sp3s* tight binding model was discussed by Nestoklon et al.  \cite{Nestoklon}. In this early work, besides the intrinsic limitation of the sp3s* model for precision modeling, effects of large strain of interface bonds were not discussed, and simply "renormalized" in the interfacial s-p two-center integrals considered as adjustable parameters. Proper consideration and full modeling of elastic relaxation actually reduces the number of adjustable parameters to few band offset parameters. The purpose of this paper is to re-examine the particular case of InAs-AlSb quantum wells using an advanced tight-binding scheme, integrating recent methodological progress in the treatment of strain, and delineate the range of parameters for which intrinsic interface states would exist in this system.

\begin{figure}
\includegraphics[width=0.45\textwidth]{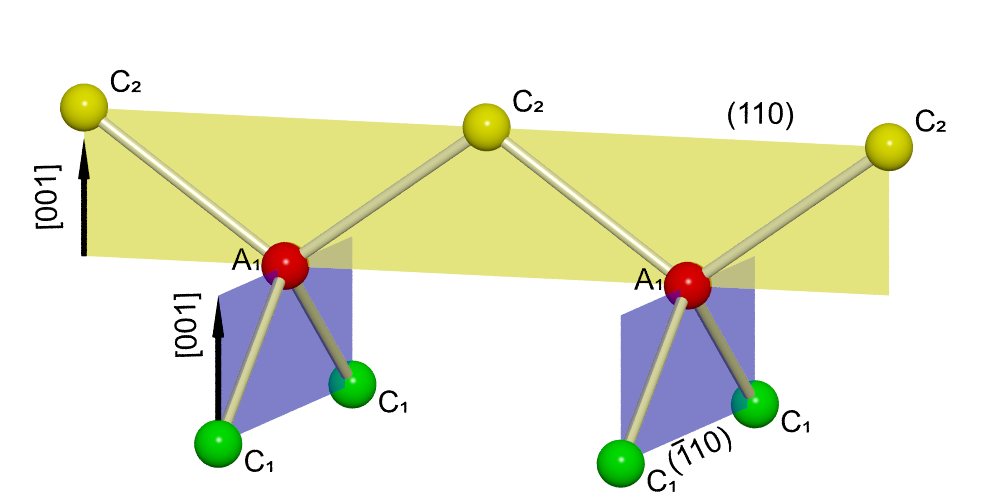}

\caption{sketch of the arrangement of chemical bonds at a $\text{C}_1\text{A}_1$/$\text{C}_2\text{A}_2$ interface  grown along the [001] direction. C and A stand for Cation and Anion species, respectively.}\label{fig:bonds}
\end{figure}

In the case of standard [001] growth axis, the mere presence of an interface between materials $\text{C}_1\text{A}_1$ and $\text{C}_2\text{A}_2$ (where A and C stand for anion and cation species) breaks not only the translational invariance but also a rotational degree of freedom, as the four-fold roto-inversion (or $S_4$) symmetry of the $T_d$ point group is no longer allowed. The atomic arrangement in the interface cell is shown in Fig.~\ref{fig:bonds} and illustrates that at the interface the $\text{C}_1\text{A}_1$ bonds lie in the $(\bar110)$ plane while the $\text{A}_1\text{C}_2$ bonds lie in the  $(110)$ plane. 

The corresponding point group symmetry is $C_{2v}$. For a symmetric quantum well with equivalent interfaces, a $S_4$ symmetry operation centered on an atom in the central layer exists and transforms one interface into the other, upgrading overall symmetry to $D_{2d}$. Although these features were clearly stated in early publications on tight-binding calculations \cite{Schulman85,Schulman851, Chang85, Smith}, it is only in the mid 90s that the resulting consequences in terms of polarization anisotropy of the optical properties were clearly observed and understood. In particular, methods for curing the native over-symmetry of classical envelope-function approach (EFA) have been proposed, following more or less explicitly the theory of invariants\cite{Krebs,Ivchenko1,Ivchenko,Nestoklon}. These methods introduce at least one new interface parameter whose value is in general not provided within the same theory and must be fixed by comparison with experiment or more elaborate calculations, therefore their predictability is limited. From this point of view, NCA interfaces are particularly problematic, because specific interface bonds exist in a single direction, either $(110)$ or $(\bar110)$, and generally undergo considerable strain. For instance, in a InAs/AlSb quantum well, the host materials are nearly lattice matched, but nominal interfaces respectively involve In-Sb bonds that are 6.3\% too long, and Al-As bonds that are 7.3\% too short. Hence, one has to cope with very large, sharply localized strain, the modeling of which requires special attention. Finally, it is worth mentioning that desired or undesired atom exchange during growth can affect the composition of the interfacial layer, so that NCA QWs can exist with either nominal $C_{2v}$ symmetry, or with same bonding at both interfaces and $D_{2d}$ symmetry, or in many intermediate, non-ideal configurations.

\section{Model}
Since atomic positions are an input of empirical tight-binding (ETB) models, the first problem to be solved is the relaxation of atomic positions under the effect of local interface strain. A zeroth order approach consists in extrapolating classical elasticity down to the single layer of chemical bonds, or molecular layer. Obviously, in order to go beyond this crude approximation, one must use atomistic elasticity such as the Valence Force Field (VFF) model \cite{ Keating, Martin}. For sake of simplicity, we shall consider that the heterostructure is strained as a whole to maintain epitaxial relation to a GaSb substrate, but adaptation to the case of a `` free-standing'' superlattice is straightforward.
In the classical elasticity limit the distance between atomic planes $i$ and $i+1$ is given, for each molecular layer, by $d_{i, i+1} = d_{i, i+1}^0 (1 + 2 c_{12}/c_{11} (a_{i, i+1}- a_s)/a_s)$ where $a_{i, i+1}$ is the lattice parameter of the pseudo-binary compound corresponding to atomic planes $i$ and $i+1$ and $a_s$ is the substrate parameter. $c_{12}$ and $c_{11}$ are the corresponding elastic constants. This result is obviously incorrect for an interface sequence like Sb=In-As since the In atom would have two highly strained ``backward'' bonds Sb=In, and two essentially unstrained ``forward'' bonds In-As.

Next we need to include equilibrium atomic positions and related strain effect in the extended-basis $spds*$ tight-binding formalism \cite{Jancu98}. For bulk materials, it is widely accepted \cite{Jancu07, Zielinski} that, in addition to changes in phase factors and power-law scaling of two center transfer integrals with interatomic distances, one should consider that the on-site orbitals (in particular, the quasi-free electron orbitals $d$ and $s*$) feel the ``geometry'' of the deformed crystal, and their energies must therefore be shifted and possibly split according to the symmetry of the deformation. It was proved that this approach leads to satisfactory fit of bulk deformation potentials. Here we use a generalization of this scheme to the situation of an atom surrounded by arbitrarily chosen partners. Say we consider a cation C surrounded with 4 different anions $\text{A}_i , i= 1-4$,  located at arbitrary positions, and need to define a local strain acting on the cation. Nominal Anion positions $\{\boldsymbol{r}_{0i}\}_{i=1\cdot\cdot4}$ are first defined, using bond lengths corresponding to $\text{CA}_i$ bulk lattice parameter and [111] bond orientations. After relaxation, this nominal, unstrained tetrahedron transforms to the actual one with atoms at positions $\{\boldsymbol{r}_{i}\}_{i=1\cdot\cdot4}$. The shapes of these tetrahedrons can be characterized using three arbitrarily chosen vectors $\{\boldsymbol{R}_{j}\}_{j=1\cdot\cdot3}$. We choose them as: $\boldsymbol{R}_1 = \boldsymbol{r}_2 - \boldsymbol{r}_1$, $\boldsymbol{R}_2 = \boldsymbol{r}_4 - \boldsymbol{r}_3$, and $\boldsymbol{R}_3 = 1/2(\boldsymbol{r}_4 + \boldsymbol{r}_3 - \boldsymbol{r}_2 -\boldsymbol{r}_1)$. It is then easy to find the matrix $T$ connecting the nominal and strained sets: $T \boldsymbol{R}_{0j} = \boldsymbol{R}_j$. The local strain tensor $\epsilon$ acting on on-site orbitals is defined by the polar decomposition $T= (1+\epsilon) R$, where $R$ is the orthogonal matrix which rotates ``nominal'' tetrahedron to the strained one. One may notice that $\epsilon$ does not fully describe local atomic configuration: It is uniquely defined by the relative coordinates of four anions surrounding given cation (or vice versa) and the change of cation position does not affect local strain tensor. To account for the cation position we introduce additional internal strain vector $\boldsymbol{u}$ defined as the (scaled to unstrained interatomic distance) displacement of cation from the centre of sphere which touches surrounding anions. Note that in a bulk material, strain and internal displacement are proportional and related by the Kleinman parameter $\zeta$, which is not the case for atomic positions in a situation of arbitrary chemical surrounding. As done implicitly in Ref.~\onlinecite{Jancu07}, we assume that the effect of the internal strain on tight-binding Hamiltonian is the same as that of the strain tensor part which transforms as a vector. In summary, the local strain hamiltonian acting on  $p$ orbitals ($p_x$, $p_y$, $p_z$) and 
$d$ orbitals with the symmetry $\Gamma_{15}$ ($d_{yz}$, $d_{zx}$, $d_{xy}$) on-site orbitals is written as:
\begin{equation}\label{ham_str_n_int}
\delta\hat{H}=
\left(\begin{array}{ccc}
\lambda_1(\sqrt3\varepsilon_1-\varepsilon_2)
   &\lambda_2(\varepsilon_{xy} + \xi u_z)
       &\lambda_2(\varepsilon_{zx} + \xi u_y)\\
\lambda_2(\varepsilon_{xy}+ \xi u_z)
   &-\lambda_1(\sqrt3\varepsilon_1+\varepsilon_2)
       &\lambda_2(\varepsilon_{yz} + \xi u_x)\\
\lambda_2(\varepsilon_{zx}+ \xi u_y)
   &\lambda_2(\varepsilon_{yz}+ \xi u_x)
       &2\lambda_1\varepsilon_2
\end{array}\right) \nonumber
\end{equation}
where we use $\varepsilon_1 = \sqrt{3}(\varepsilon_{xx}-\varepsilon_{yy})$,
$\varepsilon_2=2\varepsilon_{zz}-\varepsilon_{xx}-\varepsilon_{yy}$.
In notation of Ref.~\onlinecite{Jancu07}, for $p$-orbitals $\lambda_1=\frac12E_p\pi_{001}$, 
$\lambda_2=\frac83E_p\pi_{111}$ and for $d$-orbitals  $\lambda_1=\frac12E_d\delta_{001}$, 
$\lambda_2=\frac83E_d\delta_{111}$. The parameter $\xi$ is equal to $+1$ for cations and to $-1$ for anions. Note that in principle, $\pi$ and $\delta$ parameters may have different values for anions and cations.
These prescriptions give the same result as in Ref.~\onlinecite{Jancu07} for strained bulk semiconductors. 

However, an obvious difficulty with this general frame is the large number of parameters that need to be determined: it is clear that the sole consideration of deformation potentials at the zone center (that are reasonably well documented) does not provide enough information. On the other hand, even the most sophisticated ab initio schemes still encounter difficulties with conduction band dispersions, and blind fitting may lead to unsatisfactory parameterization. For the present purpose, one can rely on any parameterization that gives sound values of $a_{c,v}$, $b$ and $d$. Here, we reworked the strain parameters using an optimization algorithm and looking for a set of strain parameters close to the one of Ref.~\onlinecite{Jancu98} and that reproduces the recommended values of deformation potentials in the center of Brillouin zone given in Ref.~\onlinecite{Meyer}. For simplicity, since $\pi$ and $\delta$ parameters have the same effects on deformation potentials in $\Gamma$ point, $\delta$ parameters were set to zero and thus renolmalised in $\pi$ parameters.
%Here, we start with slightly reworked version of the classical set of strain parameters of Ref.~\onlinecite{Jancu98}, and just introduce one additional parameter, which is the splitting of $p$ orbitals under (111) strain, $\pi_{111}$. Indeed, the valence band of III-V compounds has a dominant p-type character, and a small splitting of $p$ orbitals efficiently splits the valence band maximum. In this approximation, hydrostatic shift of on-site energies are renormalised in the bond-length dependences of two-center integrals, and $\pi_{001}$ in the value of $\delta_{001}$. 
Strain parameters used in present calculations and resulting zone-center deformation potentials are listed in tables~\ref{tab:tab1} and~\ref{tab:tab2}.

\begin{table}
\caption{\label{tab:tab1}Strain parameters used in calculations, and resulting zone-center deformation potentials. Notations from Ref.~\onlinecite{Jancu07}}
\begin{tabular*}{\columnwidth}{@{\extracolsep{\fill}}lllll}

\hline\hline
                    & AlAs         & InAs     & AlSb       & InSb\\
\hline
$     n_s         $& $  0.6665$& $  0.7300$& $  1.0570$& $  0.4390$ \\
$     n_p         $& $  0.9626$& $  1.6282$& $  1.8518$& $  1.7595$ \\
$     n_d         $& $  1.7296$& $  1.6756$& $  1.9764$& $  1.8372$ \\
$     n_{s^*}     $& $  2.0000$& $  2.0000$& $  2.0000$& $  2.0000$ \\
$n_{ss\sigma}     $& $  2.1300$& $  2.7400$& $  2.7980$& $  4.7900$ \\
$n_{sp\sigma}     $& $  3.8720$& $  2.9100$& $  3.3320$& $  3.8540$ \\
$n_{sd\sigma}     $& $  2.1740$& $  1.4860$& $  2.1760$& $  1.7060$ \\
$n_{ss^*\sigma}   $& $  0.0000$& $  0.0000$& $  0.0000$& $  0.0000$ \\
$n_{s^*s^*\sigma} $& $  0.0000$& $  0.0000$& $  0.0000$& $  0.0000$ \\
$n_{s^*p\sigma}   $& $  1.9200$& $  1.1060$& $  1.3360$& $  1.3540$ \\
$n_{s^*d\sigma}   $& $  2.0000$& $  2.0000$& $  2.0000$& $  2.0000$ \\
$n_{pp\sigma}     $& $  2.7720$& $  2.7180$& $  2.9080$& $  1.8360$ \\
$n_{pp\pi}        $& $  4.3580$& $  4.8720$& $  3.4060$& $  4.3100$ \\
$n_{pd\sigma}     $& $  2.3330$& $  2.6650$& $  2.6710$& $  2.0050$ \\
$n_{pd\pi}        $& $  2.4729$& $  1.3749$& $  0.8629$& $  2.0769$ \\
$n_{dd\sigma}     $& $  2.6120$& $  1.3780$& $  1.1780$& $  2.9980$ \\
$n_{dd\pi}        $& $  1.7320$& $  2.7980$& $  2.9540$& $  2.9580$ \\
$n_{dd\delta}     $& $  2.8720$& $  1.0000$& $  1.6800$& $  1.0160$ \\
$\pi_{001}        $& $  0.1530$& $  0.1610$& $  0.2970$& $  0.3000$ \\
$\pi_{111}        $& $  0.3540$& $  0.4910$& $  0.6020$& $  0.3350$ \\
$\delta_{001}     $& $  0.0000$& $  0.0000$& $  0.0000$& $  0.0000$ \\
$\delta_{111}     $& $  0.0000$& $  0.0000$& $  0.0000$& $  0.0000$ \\

\hline\hline

\end{tabular*}
\end{table}

\begin{table}
\caption{\label{tab:tab2}Tight-binding deformation potentials of bulk materials. The fitted values are identical to the target values taken from Ref~\onlinecite{Meyer}, except that we use the sign convention for $a_v$, such that bandgap is proportional to $a_c-a_v$. }
\begin{tabular*}{\columnwidth}{@{\extracolsep{\fill}}lrrrr}

\hline\hline
                    & AlAs         & InAs     & AlSb       & InSb\\
\hline
$a_c $ &  $-5.64$  & $-5.08 $   &   $-4.50$   & $ -6.94$ \\
$a_v $ &  $2.47$   &  $1.00 $   &   $1.40 $   & $ 0.36 $\\
$b$    &  $-2.30$   & $-1.80$   &   $-1.35$   & $ -2.00$ \\
$d$    &  $-3.40$   & $-3.60$   &   $-4.30$   & $ -4.70$ \\
\hline\hline

\end{tabular*}
\end{table}

\section{Results and discussions}

The approach outlined in previous section is similar to that introduced by  C. Pryor \cite{Pryor} and used by R. Magri et al. for the AEPP approach \cite{Magri,Magri_InAs}, or to that used by M. Zielinski for EPTB \cite{Zielinski}. It differs however on a very important item that is the unambiguous definition of the internal displacement vector that is mandatory for proper account of trigonal deformations \cite{Jancu07}. Yet, to the best of our knowledge, these atomistic models were not validated by a crucial comparison with a well-established experimental result. A possible  test case is an In monolayer inserted in a GaAs matrix. The experimental gap is well documented with a low temperature value of 1.434 eV from optical properties of samples containing a slightly sub-monolayer amount of In giving raise to large, monolayer thick islands \cite{Brandt}. 
Our calculations give a value of 1.436 eV when using a ``natural valence band offset'' (VBO) of 0.23 eV, which is consistent with experimental VBO value for $\text{In}_{0.15}\text{Ga}_{0.85}\text{As}/\text{GaAs}$ superlattices\cite{Soucail}. Note that including the piezoelectric potential into the calculation reduces the gap by 4 meV.

Next we come to InAs/AlSb superlattices. In Table \ref{tab:1}, we compare the atomic distances in the interface regions of a 8/8 InAs/AlSb superlattice, obtained in the ``classical'' and ``atomistic'' elasticity models. For the latter, we use the Keating parameters of Ref.~\onlinecite{Keating}.
It can be observed that in the VFF calculations, interface strain perturbs the atomic positions on typically one monolayer (3\AA) on both sides of the ``anomalous'' interface bond, with an oscillatory behavior before the interplane distances stabilize to the classical elasticity value. Somewhat counter-intuitively, the deformation of anomalous interface bonds along the growth axis is larger (typically 10\%) in the VFF calculation, compared to classical elasticity.

\begin{table}
\caption{Interplane distances in the vicinity of interfaces. VFF calculations are made for a several period 8/8 InAs/AlSb superlattice grown on a GaSb substrate.\label{tab:1}         }
\begin{tabular*}{\columnwidth}{@{\extracolsep{\fill}}lccc}
\hline\hline
 & $h$ bulk & $h$ classical & $h$ VFF \\
\hline
Al-Sb & 1,5339 & 1.5437 & 1.5436 \\
Sb=Al & 1,5339 & 1.5437 & 1.5441 \\
Al-Sb & 1,5339 & 1.5437 & 1.5367 \\
Sb=In & 1,6198 & 1.7234 & 1.7323 \\
In-As & 1,5146 & 1.5041 & 1.4962 \\
As=In & 1,5146 & 1.5041 & 1.5046 \\
In-As & 1,5146 & 1.5041 & 1.5041 \\
\hline
In-As & 1,5146 & 1.5041 & 1.5041 \\
As=In & 1,5146 & 1.5041 & 1.5035 \\
In-As & 1,5146 & 1.5041 & 1.5132 \\
As=Al & 1,4150 & 1.3219 & 1.2958 \\
Al-Sb & 1,5339 & 1.5437 & 1.5541 \\
Sb=Al & 1,5339 & 1.5437 & 1.5428 \\
Al-Sb & 1,5339 & 1.5437 & 1.5436 \\
\hline\hline
\end{tabular*}
\end{table}

In table \ref{tab:2}, we show the local strain tensor associated with the different atomic sites in a 8/8 InAs/AlSb superlattice, using the VFF atomic positions. A remarkable, perhaps counterintuitive feature is the existence of a trigonal (shear) component $\epsilon_{xy}$ for atoms that have an asymmetrical chemical surrounding.

\begin{table}
\caption{local strain tensor acting on atomic orbitals on-site energies for a 8/8 InAs/AlSb superlattice grown lattice-matched to a GaSb substrate.\label{tab:2}         }
\begin{tabular*}{\columnwidth}{@{\extracolsep{\fill}}lccccccc}
\hline\hline

& $\epsilon_{xx}$(\%) & $\epsilon_{yy}$(\%) & $\epsilon_{zz}$(\%) & $\epsilon_{yz}$(\%) & $\epsilon_{zx}$(\%) & $\epsilon_{xy}$(\%) & $u_z$(\%) \\
\hline

In & -2.647 & -2.647 & 3.005 & 0. & 0. & 3.272& 4.349\\
As & 0.626 & 0.626 & -0.930 & 0. & 0. & 0. & -0.160\\
In & 0.626 & 0.626 & -0.669 & 0. & 0. & 0. &0.010\\
As & 0.626 & 0.626 & -0.686 & 0. & 0. & 0. &-0.001\\
In & 0.626 & 0.626 & -0.685 & 0. & 0. & 0. &0.\\
As & 0.626 & 0.626 & -0.685 & 0. & 0. & 0. &0.\\
In & 0.626 & 0.626 & -0.685 & 0. & 0. & 0. &0.\\
As & 0.626 & 0.626 & -0.685 & 0. & 0. & 0. &0.\\
In & 0.626 & 0.626 & -0.685 & 0. & 0. & 0. &0.\\
As & 0.626 & 0.626 & -0.685 & 0. & 0. & 0. &0.\\
In & 0.626 & 0.626 & -0.685 & 0. & 0. & 0. &0.\\
As & 0.626 & 0.626 & -0.685 & 0. & 0. & 0. &0. \\
In & 0.626 & 0.626 & -0.684 & 0. & 0. & 0. &-0.001\\
As & 0.626 & 0.626 & -0.703 & 0. & 0. & 0. &0.012\\
In & 0.626 & 0.626 & -0.403 & 0. & 0. & 0. & -0.189\\
As & 4.163 & 4.163 & -4.112 & 0. & 0. & -3.538 & 4.399\\
Al & 3.528 & 3.528 & -3.358 & 0. & 0. & -4.173 &-5.170\\
Sb & -0.645 & -0.645 & 0.950 & 0. & 0. & 0. &0.216\\
Al & -0.645 & -0.645 & 0.611 & 0. & 0. & 0. &-0.016\\
Sb & -0.645 & -0.645 & 0.636 & 0. & 0. & 0. &0.001\\
Al & -0.645 & -0.645 & 0.635 & 0. & 0. & 0. &0.\\
Sb & -0.645 & -0.645 & 0.635 & 0. & 0. & 0. &0.\\
Al & -0.645 & -0.645 & 0.635 & 0. & 0. & 0. &0.\\
Sb & -0.645 & -0.645 & 0.635 & 0. & 0. & 0. &0.\\
Al & -0.645 & -0.645 & 0.635 & 0. & 0. & 0. &0.\\

Sb & -0.645 & -0.645 & 0.635 & 0. & 0. & 0. &0.\\
Al & -0.645 & -0.645 & 0.635 & 0. & 0. & 0. &0.\\
Sb & -0.645 & -0.645 & 0.635 & 0. & 0. & 0. &0.\\
Al & -0.645 & -0.645 & 0.634 & 0. & 0. & 0. &0.001\\
Sb & -0.645 & -0.645 & 0.650 & 0. & 0. & 0. &-0.010\\
Al & -0.645 & -0.645 & 0.427 & 0. & 0. & 0. &0.139\\
Sb & -3.282 & -3.282 & 3.657 & 0. & 0. & 2.637 &-3.580\\

\hline\hline
\end{tabular*}
\end{table}

\begin{figure}
\includegraphics[width=0.45\textwidth]{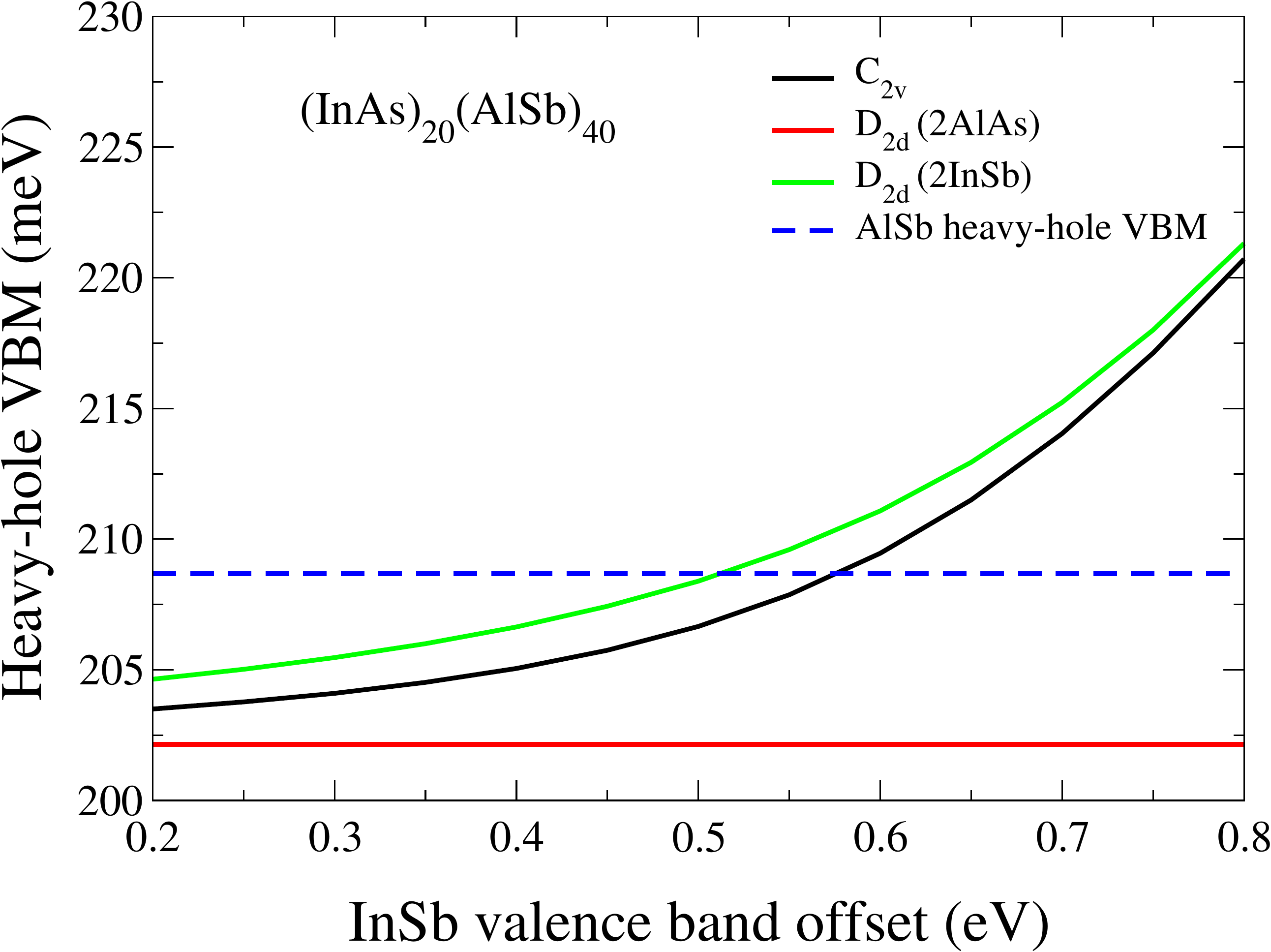}
\caption{Heavy-hole ground state energy for a 20/40  InAs/AlSb superlattice for $C_{2v}$, $D_{2d}$, 2AlAs and $D_{2d}$, 2 InSb interface configurations. The origin of energies is the unstrained InAs valence band maximum (VBM). The AlSb heavy-hole VBM is shown as the horizontal line at 208.7 meV.}\label{fig:gs}
\end{figure}

The bulk material parameters used in this work are listed in Table \ref{tab:tab3}. The calculation also requires band offset values. For the well-documented offsets for the nearly unstrained heteropairs InAs/GaSb, InAs/AlSb and GaSb/AlSb we take respectively 570, 200 and 350 meV. Experimental values agree with ab initio calculations. Unfortunately, the situation for InSb/InAs is not as clear, with no direct experimental result and a strong dispersion of theoretical predictions from 400 meV \cite{ Zunger} to 700 meV \cite{Van_der_Walle}. 

\begin{figure}
\includegraphics[width=0.45\textwidth]{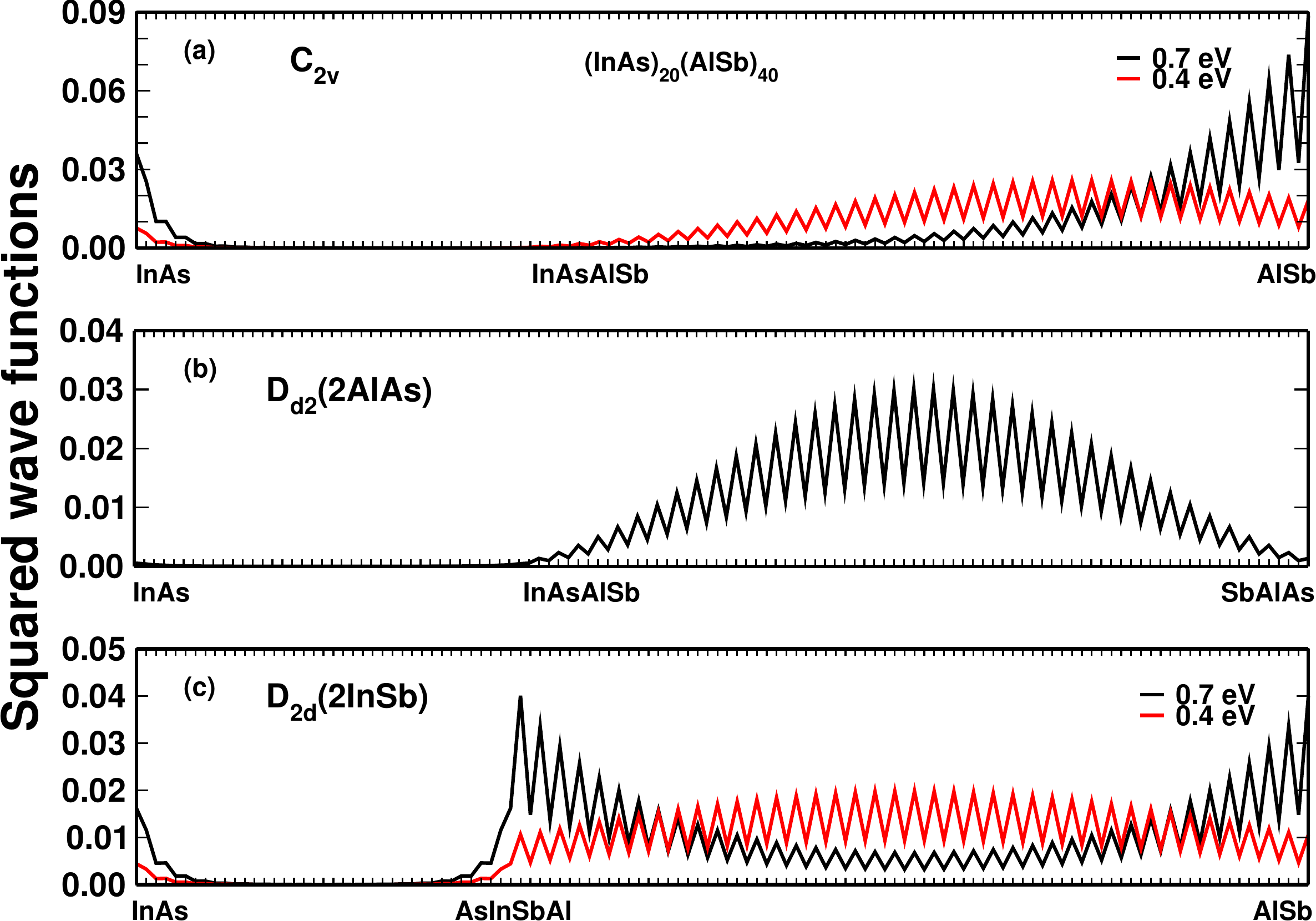}
\caption{a)Heavy hole ground state wave function for a 20/40 InAs/AlSb SL, for two values of the InSb band offset, 700 and 400 meV. A solid line is drawn to connect tight binding on-site amplitudes for clarity. In all these materials, valence band eigenstates have a strongly dominant Anion character; b, c) Same for the $D_{2d}$, 2 AlAs and $D_{2d}$, 2 InSb configurations.}\label{fig:wf}         
\end{figure}

The molecular layer of InSb certainly act as a ``potential well'' in the valence band, but uncertainty in VBO implies that the depth of this trap is unknown, and we shall consider it as the only free parameter in the calculations. Conversely, the AlAs bonds act as a ``potential barrier'' whose height is also not so well documented, but the impact of this uncertainty on our results is actually negligible. 
When epitaxial strain is taken into account, the AlSb heavy-hole band extremum lies at 208.7 meV above the arbitrary reference level of unstrained InAs VBM. The heavy hole confinement in a regular 12.2 nm-thick (40 monolayers) QW is 6.4 meV. Hence, one reasonably expects a hole ground state at 202.3 meV. Fig.~\ref{fig:gs} shows the evolution of actual hole ground state energy in a 20/40 InAs/AlSb superlattice as a function of InSb band offset, for the 3 ideal situations: ``$C_{2v}$'' (AlAs and InSb interfaces), ``$D_{2d}$'' (2 AlAs) and ``$D_{2d}$'' (2 InSb). 
In $C_{2v}$ case for InAs/InSb VBO larger than 600 meV, the ground state is above the AlSb VBM, hence it is clearly trapped by the InSb bond. For smaller offsets, the ground state lies between the AlSb VBM and 204 meV: the situation is better described as a regular quantum well with an attractive perturbation at one interface, that decreases the confinement energy and polarizes the wavefunction. As shown in Fig.~\ref{fig:wf}a, in all cases the wavefunction is strongly asymmetrical with respect to the center of AlSb layer: As long as AlSb thickness remains finite, it is difficult to define a rigorous criterium for existence of an interface state. By exploring numerically larger layer thicknesses, we find that 500 meV is a practical threshold offset value for the existence of an interface state at a InSb interface between InAs and AlSb.

\begin{table}
\caption{\label{tab:tab3}Tight-binding parameters used in calculations.}
\begin{tabular*}{\columnwidth}{@{\extracolsep{\fill}}lrrrr}

\hline\hline
                    & InAs         & InSb     & AlAs       & AlSb\\
\hline
$              a$   & $    6.0580$ & $ 6.4794$ & $ 5.6600$ & $ 6.1355$ \\
$          E_{s}^a$ & $   -6.4738$ & $-6.1516$ & $-5.9874$ & $-6.0025$ \\
$        E_{s^*}^a$ & $   16.8502$ & $14.7582$ & $19.5074$ & $16.4623$ \\
$          E_{s}^c$ & $   -0.1418$ & $-0.3634$ & $ 0.9483$ & $ 0.6705$ \\
$        E_{s^*}^c$ & $   16.8393$ & $14.8015$ & $19.5038$ & $16.4797$\\
$          E_{p}^a$ & $    2.4784$ & $ 2.1150$ & $ 3.4914$ & $ 2.5476$\\
$          E_{d}^a$ & $   11.3833$ & $ 9.8811$ & $13.0560$ & $11.1777  $\\
$          E_{p}^c$ & $    5.2829$ & $ 5.5198$ & $ 6.3335$ & $ 5.8536$\\
$          E_{d}^c$ & $   11.3991$ & $ 9.9511$ & $13.0592$ & $11.1500$\\
$         ss\sigma$ & $   -1.5096$ & $-1.2228$ & $-1.8436$ & $-1.4804$\\
$   s_as^*_c\sigma$ & $   -2.0155$ & $-1.6619$ & $-1.7884$ & $-2.9492$\\
$   s^*_as_c\sigma$ & $   -1.1496$ & $-1.3929$ & $-1.3059$ & $-1.4096$\\
$     s^*s^*\sigma$ & $   -3.3608$ & $-2.8985$ & $-3.6128$ & $-1.2369$\\
$     s_ap_c\sigma$ & $    2.2807$ & $ 2.2046$ & $ 2.5778$ & $ 2.2550$\\
$     s_cp_a\sigma$ & $    2.6040$ & $ 2.3639$ & $ 2.7962$ & $ 2.5961$\\
$   s^*_ap_c\sigma$ & $    1.9930$ & $ 1.6962$ & $ 2.1581$ & $ 2.1314$\\
$   s^*_cp_a\sigma$ & $    2.0708$ & $ 1.9879$ & $ 2.2397$ & $ 1.9456$\\
$     s_ad_c\sigma$ & $   -2.8945$ & $-2.3737$ & $-2.5623$ & $-2.5320$\\
$     s_cd_a\sigma$ & $   -2.3175$ & $-2.1766$ & $-2.3841$ & $-2.0483$\\
$   s^*_ad_c\sigma$ & $   -0.6393$ & $-0.5548$ & $-0.8045$ & $-0.5304$\\
$   s^*_cd_a\sigma$ & $   -0.5949$ & $-0.4875$ & $-0.7491$ & $-0.5989$\\
$         pp\sigma$ & $    3.6327$ & $ 3.4603$ & $ 4.1970$ & $ 3.7007$\\
$            pp\pi$ & $   -0.9522$ & $-1.1630$ & $-1.3145$ & $-1.2989$\\
$     p_ad_c\sigma$ & $   -1.1156$ & $-1.3928$ & $-1.6473$ & $-1.3211$\\
$     p_cd_a\sigma$ & $   -1.3426$ & $-1.4145$ & $-1.7603$ & $-1.7320$\\
$        p_ad_c\pi$ & $    1.2101$ & $ 1.1921$ & $ 1.7646$ & $ 1.6944$\\
$        p_cd_a\pi$ & $    1.5282$ & $ 1.7536$ & $ 2.1099$ & $ 1.7783$\\
$         dd\sigma$ & $   -0.8381$ & $-0.6688$ & $-1.2241$ & $-0.9481$\\
$            dd\pi$ & $    1.9105$ & $ 1.4601$ & $ 2.1770$ & $ 1.8128$\\
$         dd\delta$ & $   -1.3348$ & $-1.4373$ & $-1.7585$ & $-1.6148$\\
$\Delta_a/3$        & $    0.1558$ & $ 0.3810$ & $ 0.1721$ & $0.3454$\\
$\Delta_c/3$        & $    0.1143$ & $ 0.1275$ & $ 0.0072$ & $0.0121$\\ 
\hline\hline

\end{tabular*}
\end{table}

\begin{figure}
\includegraphics[width=0.45\textwidth]{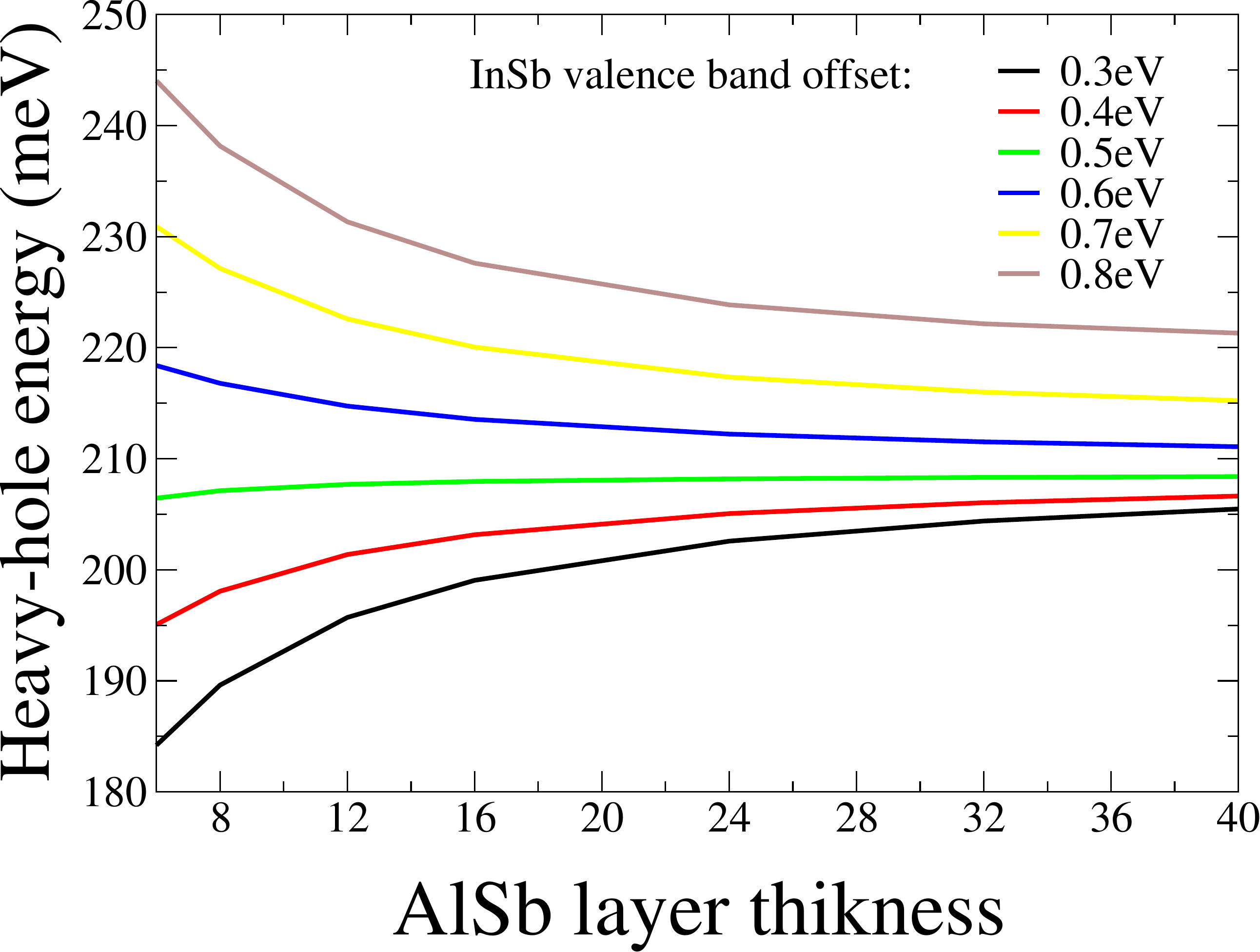}
\caption{Dependence of heavy-hole ground state energy on AlSb layer thickness for the $D_{2d}$, 2 InSb configuration.}\label{fig:gs_width}
\end{figure}

\begin{figure}
\includegraphics[width=0.45\textwidth]{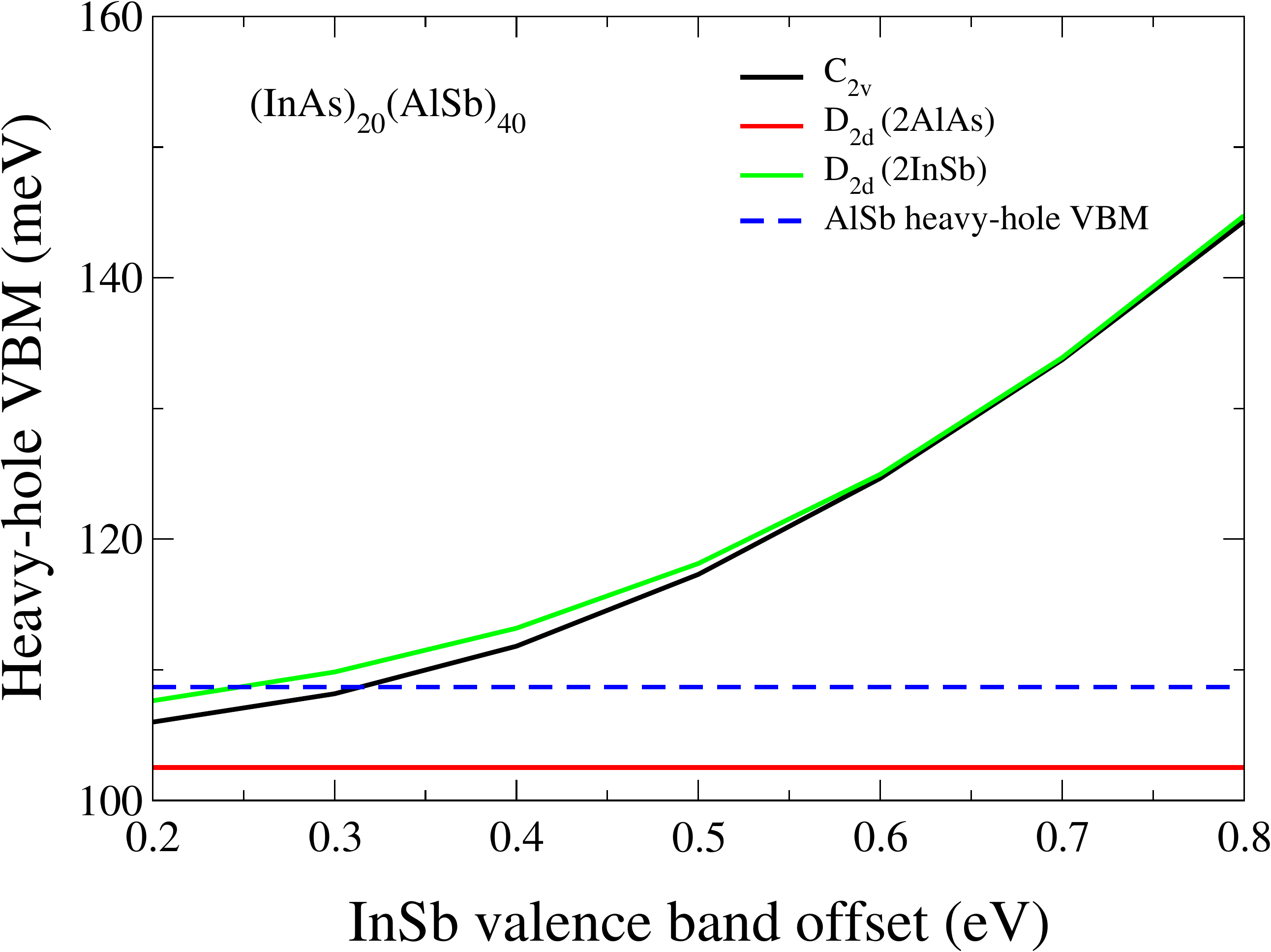}
\caption{Analogous to Fig~\ref{fig:gs} with InAs/AlSb VBO~=~100~meV.}\label{fig:gs1}
\end{figure}

The $D_{2d}$ case with two AlAs interfaces (see Figs.~\ref{fig:gs}, \ref{fig:wf}b) corresponds to the regular quantum well case. The $D_{2d}$ 2 InSb configuration is more interesting, because interface states may exist at both interfaces, and combine into symmetric (bonding) or antisymmetric (antibonding) states, with a splitting depending strongly on AlSb layer thickness. Due to this interaction, the ground state energy remains nearly constant when decreasing the AlSb thickness. This result is illustrated in Fig.~\ref{fig:gs_width}. A similar trend is also valid for interface state coupling through the InAs layer, but to a much smaller extent due to the fast decay of interface state into InAs. Note however that  his remark holds only in as much as the unavoidable difference between the two interface state energies is smaller than their mutual coupling. Results displayed in Fig.~\ref{fig:wf} are qualitatively similar to those obtained from LDA calculations by Shaw et al. \cite{Shawetal}, but the decay of interface state in AlSb is much slower in our calculations. It is noteworthy that calculations of Ref.~\onlinecite{Shawetal} do not include spin-orbit interaction and therefore, the effective valence band quantum wells differ. Finally, it is interesting to examine the effect of a change in InAs/AlSb VBO, within the experimental uncertainty range $150\pm50$meV: the smaller this offset is, the easier it is for the interface trap to capture the valence ground state. For vanishing InAs/AlSb VBO, any value of InSb VBO would lead to an interface state, whose decay in both layers would become almost symmetrical. In Fig~\ref{fig:gs1}, we show the dependencies on InSb/InAs offset for a InAs/AlSb offset of 100 meV. It is seen that in this case, the interface state exists as soon as the InSb/InAs offset exceeds 314 meV.

So far, we have discussed interface potential in terms of a ``diagonal'' or scalar contribution. However, as mentioned in the introduction, interface also break a rotational invariance and corresponding Hamiltonian admixes heavy and light holes, which results in the linear polarization of optical spectra when system has $C_{2v}$ symmetry\cite{Ivchenko,Krebs,Theodorou}. Here, valence states are confined in the AlSb layer, and ground state peaks close to the InSb bonds. They undergo the strong spin-orbit coupling of Sb. Since spin-orbit interaction tends to force total angular momentum eigenstates, weak polarization anisotropy is expected. This anisotropy is confirmed by the calculations: For the 20/40 InAs/AlSb superlattice, we obtain a degree of linear polarization of the ground optical transition (with principal axis along the [1,1,0] and [-1,1,0] directions) equal to 6\%.

\section{Conclusion}
We have used extended-basis tight-binding to model the no-common atom system of InAs/AlSb with the highest possible accuracy. We find that for a large range of the natural band offset of InSb, there exists an intrinsic interface state ``trapped'' by the plane of interfacial InSb bonds. The existence of such a state is important for valence band physics in this system, but also plays an important role in the material characterization using interband optics.

\section{Acknowledgements}
The authors thank Dr. F. Glas for fruitful discussions and for cross-checking VFF atomic positions. This work was supported in part by ANR "SNAP", by French-Russian Int. Laboratory "ILNACS", and by "Triangle de la Physique CAAS".

\bibliography{InAs_AlSb}
\bibliographystyle{apsrev4-1}

\end{document}